\documentclass[twocolumn,showpacs,preprintnumbers,amsmath,amssymb]{revtex4}

\usepackage{graphicx}
\usepackage{dcolumn}
\usepackage{bm}
\usepackage{xcolor}

\begin{document}

\preprint{APS/123-QED}

\title{Experimental benchmark of the quantum–classical crossover in a spin ladder}


\author{Hironori Yamaguchi$^{1,2}$, Itsuki Shimamura$^{1}$, Akira Matsuo$^{3}$,\\ Koichi Kindo$^{3}$, Koji Araki$^{4}$, Yoshiki Iwasaki$^{2,5}$, and Masayuki Hagiwara$^{6}$}
\affiliation{
$^1$Department of Physics, Osaka Metropolitan University, Osaka 558-8585, Japan\\
$^2$Innovative Quantum Material Center (IQMC), Osaka Metropolitan University, Osaka 558-8585, Japan\\
$^3$Institute for Solid State Physics, the University of Tokyo, Chiba 277-8581, Japan\\
$^4$Department of Applied Physics, National Defense Academy, Kanagawa 239-8686, Japan\\
$^5$Department of Physics, Meisei University, Tokyo 191-8506, Japan
$^6$Center for Advanced High Magnetic Field Science (AHMF), Graduate School of Science, the University of Osaka, Osaka 560-0043, Japan
}

Second institution and/or address\\
This line break forced

\date{\today}

\begin{abstract} 
We report a spin-(1/2, 5/2) three-leg ladder realized in a radical–Mn polymer, exhibiting an antiferromagnetic transition and magnetization curves accurately described by classical mean-field theory. 
Although the underlying spin model intrinsically supports strong quantum fluctuations, as confirmed by quantum Monte Carlo simulations, the real system shows an anomalously complete suppression of quantum behavior. 
These findings provide a key experimental benchmark for the quantum–classical crossover and suggest that lattice topology can play a crucial role in tuning the balance between quantum and classical physics in strongly correlated systems.
\end{abstract}

\maketitle 
The interplay between quantum fluctuations and classical correlations in strongly correlated systems is a central theme of contemporary condensed matter physics. 
The so-called quantum–classical crossover or criticality has been recognized as a fertile ground for emergent phenomena, including high-temperature superconductivity, non-Fermi-liquid behavior, and unconventional magnetism~\cite{sp1,sp2,frust}. 
In these systems, the competition between classical order tendencies and quantum fluctuations gives rise to unconventional ground states and anomalous responses, often with no classical counterpart.
One-dimensional (1D) quantum spin systems are a paradigmatic example of how quantum fluctuations can dominate and destabilize classical magnetic order~\cite{TLL1, TLL2, TLL3, Haldane}. 
In contrast, large-spin systems and higher-dimensional lattices are generally expected to exhibit robust classical order with negligible quantum effects. 
Understanding how quantum and classical behaviors coexist and compete, particularly in systems with mixed spin degrees of freedom and nontrivial lattice topologies, is therefore an important challenge in the search for novel quantum phases and control principles.

Among various platforms for exploring the interplay between quantum and classical behaviors, spin ladders have emerged as particularly attractive representatives of quantum spin systems~\cite{ladder1,SL_exp2,SL_exp3,SL_exp4,3Cl4FV}.
These systems not only exhibit strong quantum fluctuations intrinsic to one-dimensionality, but also possess a unique lattice topology that allows for tunable ground-state properties depending on the number of legs.
Notably, spin-1/2 antiferromagnetic (AF) spin ladders demonstrate a striking dichotomy: even-leg ladders realize a singlet ground state with a finite excitation gap, whereas odd-leg ladders host gapless excitations, as confirmed both theoretically and experimentally~\cite{ladder2,ladder3}.
This sensitivity of the ground state to the ladder topology exemplifies the profound quantum nature of these systems.
Moreover, spin ladders can be viewed as intermediate between 1D chains and two-dimensional (2D) square lattices.
Even though the 2D Heisenberg square lattice retains some quantum fluctuation effects on magnetic moments and excitations, its ground state exhibits long-range AF order~\cite{square}, and it is well known as the parent structure of high-$T_{\rm{c}}$ cuprates~\cite{highTc1,highTc2}.
Increasing the number of legs effectively interpolates between these limits, making spin ladders an ideal stage for investigating the emergence and suppression of quantum behavior as classical correlations become increasingly dominant.
In this context, spin ladders offer a unique opportunity to study quantum–classical crossover phenomena under well-controlled conditions with clear theoretical interpretation.

In this paper, we realize a spin-$(1/2,5/2)$ three-leg ladder in a radical-Mn polymer [Mn($p$-Py-V)$_2$(NO$_3$)$_2$]$_n$ ($p$-Py-V = 3-(4-pyridinyl)-1,5-diphenylverdazyl,). 
Magnetic susceptibility and specific heat measurements indicated a phase transition to an AF ordered state.
We explain the electron spin resonance (ESR) modes and the magnetization curve in the ordered phase using a mean-field (MF) approximation.
Furthermore, while the model inherently supports quantum fluctuations as shown by quantum Monte Carlo simulations, the real system exhibits an anomalously complete suppression of quantum behavior, highlighting the striking discrepancy between theory and experiment.

The X-ray intensity data were collected using a Rigaku XtaLAB Synergy-S instrument.
The crystal structures displayed in this paper were drawn using VESTA software~\cite{vesta}.
Magnetization were measured using a commercial SQUID magnetometer (MPMS, Quantum Design).
High-field magnetization in pulsed magnetic fields was measured using a non-destructive pulse magnet.
Specific heat measurements were performed using a commercial calorimeter (PPMS, Quantum Design). 
All experiments were performed using powder samples.
The molecular orbital (MO) calculations were performed using the UB3LYP method and a basis set of 6-31G~\cite{MOcal}. 
The quantum Monte Carlo (QMC) simulation was performed for $N$ = 288 under the periodic boundary condition~\cite{QMC2,ALPS,ALPS3}. 


The crystal structure of [Mn($p$-Py-V)$_2$(NO$_3$)$_2$]$_n$ is shown in Fig. 1(a).
The crystallographic parameters at 100 K are as follows: monoclinic, space group $P2_1$, $a$ = 5.1157(2)  $\rm{\AA}$, $b$ = 13.3111(6) $\rm{\AA}$, $c$ = 27.5413(11) $\rm{\AA}$ (see Supplementary Information~\cite{supple}).
In the $p$-Py-V molecule, the central ring comprises four nitrogen (N) atoms with a maximum spin density~\cite{frust_honeycomb}, leading to a localized spin system. 
Hence, the verdazyl radicals, $p$-Py-V, and Mn$^{2+}$ ions possess spin values of 1/2 and 5/2, respectively. 
The MnO$_6$ octahedra are linked into a 1D polymeric chain through bridging nitrate ligands, as shown in Fig. 1(a). 
This arrangement gives rise to a well-defined 1D coordination polymer.
The MO calculations were used to evaluate exchange interactions attributed to MO overlaps between radicals.
We evaluated a predominant AF exchange interaction $J_{1}/k_{\rm{B}}$ = 38 K, forming a uniform spin-1/2 chain along the $a$ axis, as shown in Fig. 1(b).
Each radical spin ($\boldsymbol{s}$) in the 1D chain is coupled with the Mn spin ($\boldsymbol{S}$) via intramolecular interactions $J_{2}$. 
The exchange interaction between adjacent Mn$^{2+}$ spins, $J_{3}$, is mediated by a bridging nitrate ligand through an Mn–O–N–O–Mn pathway. 
Due to the inefficient superexchange through this indirect path, the interaction is expected to be weak, on the order of 0.1–1 K.
Consequently, a spin-(1/2,5/2) three-leg ladder is formed along the $a$ axis, as shown in Fig. 1(b).

\begin{figure}[t]
\begin{center}
\includegraphics[width=20pc]{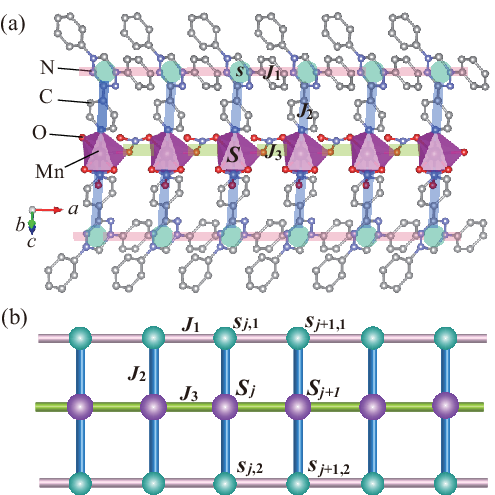}
\caption{(a) Crystal structure of [Mn($p$-Py-V)$_2$(NO$_3$)$_2$]$_n$ forming the three-leg ladder along the $a$ axis. 
Hydrogen atoms are excluded to enhance clarity. 
The green nodes represent the spin-1/2 of the radicals. 
The thick lines represent exchange interactions.
(b) Spin-(1/2,5/2) three-leg ladder comprising $J_{\rm{1}}$, $J_{\rm{2}}$, and $J_{\rm{3}}$. 
$\boldsymbol{s}$ and $\boldsymbol{S}$ denote the spins on the radical and Mn$^{2+}$, respectively. 
}
\end{center}
\end{figure}

Figures 2(a) shows the magnetic susceptibility ($\chi$ = $M/H$) at 0.1 T.
A discontinuous change was observed at approximately $T_{\rm{N}}$ = 9.0 K, which can be attributed to a phase transition to an AF ordered state owing to finite interchain interactions. 
The experimental result for the specific heat $C_{\rm{p}}$ at zero-field exhibited a peak at $T_{\rm{N}}$, demonstrating the phase transition, as shown in Fig. 2(b).
In the low-temperature region below $\sim$5 K, the $C_{\rm{p}}$ approaches a $T$ -linear behavior that is attributed to a linear dispersive mode in 1D AF systems. 
This behavior indicates that the exchange interactions forming the 1D spin ladder are indeed effective in the present compound. 
Under applied magnetic fields, $T_{\rm{N}}$ gradually shifts to lower temperatures as the field increases. 
Figures 2(c) presents the magnetization curves under a pulsed magnetic field.
We observe a nonlinear increase despite below $T_{\rm{N}}$, and it is almost independent of the temperature.
A significant bending appears at approximately 10 T, where the magnetization corresponds to about 5/7 of the saturation value. 
Considering the magnetic unit composed of two spin-1/2 sites and one spin-5/2 site, the 5/7 magnetization bending suggests nearly fully polarized state of the spin-5/2. 
In the low-field region, we observe a discontinuous change at $H_{\rm{c}}$$\sim $2.0 T, indicates a spin-flop transition caused by a small magnetic anisotropy, as shown in the inset of Fig. 2(c). 

\begin{figure}
\begin{center}
\includegraphics[width=20pc]{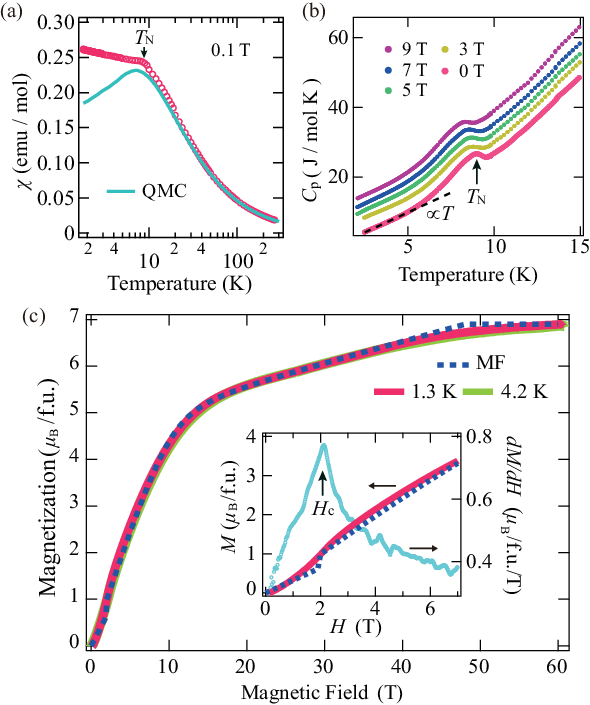}
\caption{(a) Temperature dependence of magnetic susceptibility ($\chi$ = $M/H$) of [Mn($p$-Py-V)$_2$(NO$_3$)$_2$]$_n$ at 0.1 T.
The solid line represents the QMC result for $J_{1}/k_{\rm{B}}$  = 29 K, $J_{2}/k_{\rm{B}}$  = 2.5 K, $J_{3}/k_{\rm{B}}$  = 1.1 K.
(b) Temperature dependence of the specific heat $C_{\rm{p}}$ of [Mn($p$-Py-V)$_2$(NO$_3$)$_2$]$_n$.
The arrow indicates the phase transition temperatures at zero field.
For clarity, the values for 3, 5, 7, and 9 T have been shifted up by 4, 6.5, 9, and 12 J/mol K, respectively.
(c) Magnetization curve of [Mn($p$-Py-V)$_2$(NO$_3$)$_2$]$_n$ at 1.3 and 4.2 K. 
The broken line represents the MF result with $D/k_{\rm{B}}$  = 0.1 K, averaged over field directions for the powder sample.
A radical purity of 95 ${\%}$ is considered. 
The inset shows the low-field region with $dM/dH$ at 1.3 K, and the arrow indicates the spin-flop transition at $H_{\rm{c}}$.
}\label{f2}
\end{center}
\end{figure}

\begin{figure}
\begin{center}
\includegraphics[width=20pc]{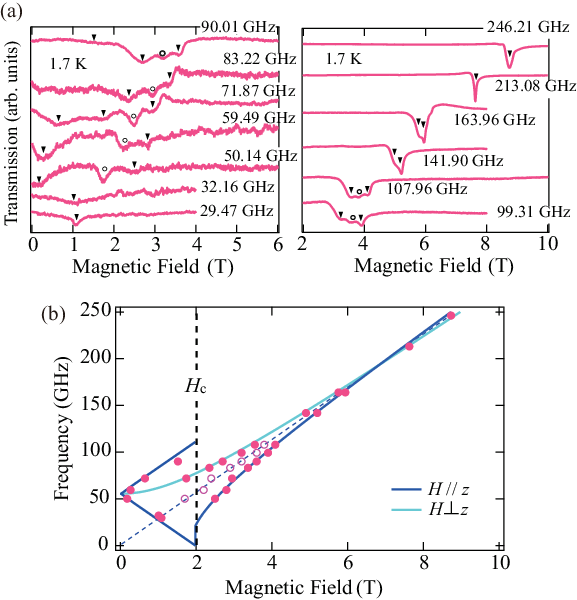}
\caption{(a) Frequency dependence of ESR absorption spectra of [Mn($p$-Py-V)$_2$(NO$_3$)$_2$]$_n$ at 1.7 K.
Closed symbol marks the AF resonance fields, while open symbol marks the paramagnetic resonance fields caused probably by impurities.
(b) Frequency-field plot of the resonance fields.
The closed and open circles correspond to the resonance fields indicated by the closed and open symbols in Fig. 3(a), respectively.
Solid and broken lines indicate the calculated easy-axis AF resonance modes and paramagnetic resonanse line, respactively.
The vertical broken line indicates the spin flop transition field $H_{\rm{c}}$.
}\label{f3}
\end{center}
\end{figure}

\begin{figure}
\begin{center}
\includegraphics[width=20pc]{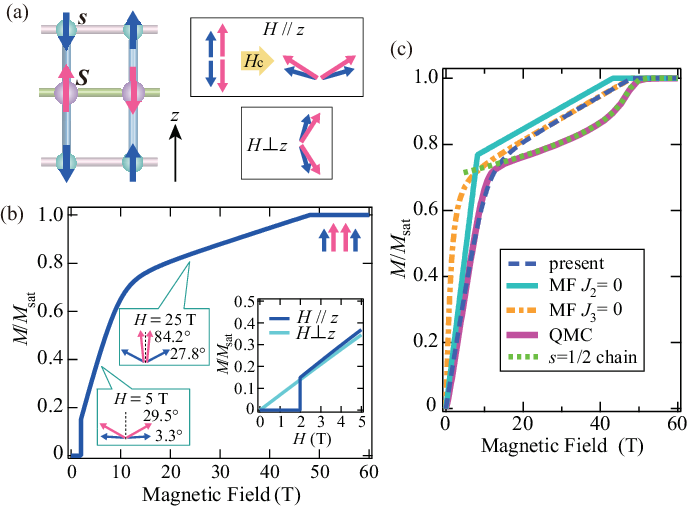}
\caption{(a) Schematic views of the spin configurations. 
The larger pink and smaller blue arrows represent the sublattices associated with $\boldsymbol{S}$ and $\boldsymbol{s}$, respectively.
The spins form a collinear AF structure at zero field, while for $H\parallel z$ with $H \textgreater H_{\rm{c}}$ and for $H\perp z$, the spins exhibit canted configurations. 
Two sublattice pairs, composed of $\boldsymbol{s}$-$\boldsymbol{s}$ and $\boldsymbol{S}$-$\boldsymbol{S}$, polarize toward the field direction with different canting angles. 
(b) Magnetization curve for $H\parallel z$, calculated using the MF approximation with $J_{1}/k_{\rm{B}}$  = 29 K, $J_{2}/k_{\rm{B}}$  = 2.5 K, $J_{3}/k_{\rm{B}}$  = 1.1 K, $D/k_{\rm{B}}$  = 0.1 K.
The illustration depicts the tilt of the sublattice pairs toward the external field, with canting angles defined as 90°for parallel alignment.
The inset shows the low-field region for $H\parallel z$ and $H\perp z$.
(c) Calculated magnetization curves of the present ladder, showing the effects of removing $J_{2}$ or $J_{3}$, and comparison between MF and QMC results, assuming $D$ =0.
The result for the $s$=1/2 chain is calculated by using QMC with $H_{\rm{in}} \approx -4.7$ T, and the magnetization is offset by 5/7. 
}\label{f3}
\end{center}
\end{figure}

The frequency dependence of the ESR absorption spectra in the ordered phase is presented in Fig. 3(a).
Because the experiments were performed using the powder samples, the observed signals corresponded to the resonance fields for the external fields parallel to the principal axes. 
The resonance signals at low frequencies exhibit a number of resonance fields and obviously deviate from the linear function of the field, while those at high frequencies are almost proportional to the external field. 
All the resonance fields are plotted in the frequency-field diagram, as shown in Fig. 3(b). 
Since a zero-field gap of $\sim$55 GHz, which corresponds to the energy scale of $H_{\rm{c}}$, is expected from the extrapolation of the resonance modes, the observed resonance fields suggest AF resonance modes with easy-axis anisotropy~\cite{Kittel,a235Cl3V,FeCl4}. 
The resonance signal observed along the line passing through the origin is attributed to impurity contributions, most likely arising from minor lattice defects.

We analyzed the observed ESR modes using a MF approximation assuming the three-leg ladder model with an easy-axis anisotropy of $\boldsymbol{S}$ (see Supplementary Information~\cite{supple}). 
Then, the spin Hamiltonian is given by
\begin{multline}
\mathcal {H} = J_{\rm{1}}{\sum^{}_{j}}{\sum^{}_{{\alpha}=1,2}}\boldsymbol{s}_{j,\alpha}{\cdot}\boldsymbol{s}_{j+1,\alpha}
+J_{\rm{2}}{\sum^{}_{j}}{\sum^{}_{{\alpha}=1,2}}\boldsymbol{s}_{j,\alpha}{\cdot}\boldsymbol{S}_{j}\\
\hspace*{\fill}
+J_{\rm{3}}{\sum^{}_{j}}\boldsymbol{S}_{j}{\cdot}\boldsymbol{S}_{j+1}
-D{\sum^{}_{j}}(S_{j}^{z})^2-g\mu_{\rm{B}}{\sum^{}_{j}}\textbf{{\textit H}}{\cdot}(\boldsymbol{s}_{j}+\boldsymbol{S}_{j})
\end{multline}
where $\textbf{{\textit H}}$ denotes the external magnetic field, $D$ is on-site anisotropy ($D$$\textgreater$0), $g$=2.0 for the organic radical and the Mn$^{2+}$ ion.
Although the easy-axis direction $z$ cannot be determined experimentally because of the use of powder samples, it is assumed here, for convenience, to be perpendicular to the chain direction. 
This assumption does not affect the analysis results. 
The spin structure is described by the four-sublattice model, as shown in Fig. 4(a).
Evaluating the free energy, we can determine stable spin configuration, leading to a typical collinear state along the easy axis ($z$ axis) at zero field.
For $H\parallel z$, the discontinuous spin-flop transition occurs at a critical field $H_c$. 
Above $H_c$, the two sublattice pairs associated with $\boldsymbol{s}$ and $\boldsymbol{S}$ are tilted toward the field direction with different canting angles. 
A similar configuration is realized under magnetic fields even for $H\perp z$.
We derive the resonance conditions by solving the equation of motion for the sublattice moments~\cite{43,44}.
Considering the experimental value of $H_c$, we determined the parameters mainly by fitting the ESR data, while taking into account the magnetization for consistency.
The best fit between the experimental and calculated results was obtained with the following parameters: $J_{1}/k_{\rm{B}}$  = 29(2) K, $J_{2}/k_{\rm{B}}$  = 2.5(5) K, $J_{3}/k_{\rm{B}}$  = 1.1(2) K, $D/k_{\rm{B}}$  = 0.10(4) K, as shown in Fig. 3(b).
The energy scale of $J_{1}$ is consistent with that evaluated from the MO calculation, and the energy scale of $D$ is the same as those of similar veradzyl-based complexes with Mn$^{2+}$ spin~\cite{Mn1,Mn2}.
Furthermore, the ESR spectra are quantitatively reproduced by the four-sublattice model based on a single three-leg ladder, indicating that interladder coupling effects do not significantly affect the observed spin dynamics.

The magnetization curve calculated by the MF approximation with the evaluated parameters are shown in Fig. 4(b).
For $H\parallel z$, a discontinuous change appears at $H_c$, whereas for $H\perp z$, the curve exhibits only a monotonic increase.
Notably, the calculated curve for $H \textgreater H_c$ displays a continuous, gradual increase despite being obtained from a classical MF model at $T=0$ without thermal or quantum fluctuations. 
This behavior reproduces the observed magnetization curve well, as shown in Fig. 2(c). 
In the canted spin configuration, below approximately 10 T, the $\boldsymbol{S}$ sublattices mainly cant toward the field direction, while the $\boldsymbol{s}$ sublattices remain almost unchanged, as illustrated in Fig 4(b). 
This behavior is attributed to the larger moment size of $\boldsymbol{S}$ and the weaker $J_{3}$, which gains more Zeeman energy when aligned with the field.
Above approximately 10 T, the $\boldsymbol{S}$ sublattices become nearly fully polarized, and the $\boldsymbol{s}$ sublattices begin to cant toward the field direction. 
The saturation field is estimated to be around 48 T, although it is not observed experimentally, presumably due to thermal fluctuations. 
For $J_{2}=0$, the $\boldsymbol{s}$ and $\boldsymbol{S}$ become decoupled and cant independently toward the field direction.
In this case, the $\boldsymbol{S}$ sublattice initially becomes fully polarized, resulting in a discontinuous change, as shown in Fig. 4(c).
For $J_{3}=0$, the AF correlation between the $\boldsymbol{S}$ spins vanishes. 
The $\boldsymbol{S}$ sublattices consequently behave almost paramagnetically, leading to a steep increase in the magnetization, as shown in Fig. 4(c).

We examine the spin state of the present three-leg ladder.
In the present system, the weak but finite easy-axis anisotropy of the $\boldsymbol{S}$ spins is expected to propagate effectively through the entire ladder via the $ J_{2}$ coupling, stabilizing the four-sublattice Néel-like ground state by breaking the SU(2) spin-rotation symmetry.
To highlight the interplay between classical correlations and quantum fluctuations in the present model, we compare the MF results with the QMC calculation that includes quantum effects. 
As shown in Fig. 2(a), the magnetic susceptibility calculated by QMC reproduces the experimental results for $T$ $\gg $ $T_{\rm N}$, where classical correlations within the ladder are not yet developed.
Upon approaching $T_{\rm N}$, however, a gradual deviation from the calculated curve appears, reflecting the development of classical short-range correlations within the ladder that precede the long-range order.
This crossover behavior naturally corresponds to the energy scale of the intra-ladder exchange interactions.
As for the magnetization curve, the QMC result exhibits a pronounced nonlinear behavior above approximately 10 T, as shown in Fig. 4(c).
In this high-field region, the $\boldsymbol{S}$ spins become almost fully polarized, effectively decoupling from the ground-state dynamics~\cite{zigzag,kondo}. 
This leads to an effective spin-1/2 AF uniform chain composed of the $\boldsymbol{s}$ spins, whose ground state is described by a Tomonaga–Luttinger liquid (TLL)~\cite{TLL1,TLL2,TLL3}.
Actually, the high-field behavior is almost identical to that of the spin-1/2 chain with an effective internal field $H_{\rm{in}}$, originating from the second term of Eq. (1), given by -$\frac{5}{2}\frac{J_{2}k_{\rm{B}}}{g\mu_{\rm{B}}} \approx -4.7$ T, as shown in Fig. 4(c).
The qualitative difference in the high-field magnetization between classical and quantum calculations is thus attributed to strong quantum fluctuations inherent in the effective spin-1/2 chain. 

In typical quantum spin systems, even when long-range ordering emerges due to finite interchain or interlayer couplings, the application of external magnetic fields gives rise to characteristic nonlinear magnetization behavior associated with quantum fluctuations and field-induced quantum phases, such as gapped plateaus and spin-liquid-like states~\cite{a235Cl3V,sankaku,BaCo,PF6}.
Notably, even in other mixed spin-(1/2,5/2) systems, quantum effects typically persist under magnetic fields~\cite{FeCl4,mix1,mix2}. 
By contrast, in the present system, the magnetization is fully reproduced by the MF approximation, clearly demonstrating that quantum fluctuations are almost completely quenched, despite the presence of effective correlations characteristic of a quantum spin chain. 
These findings reveal that the anomalous quenching of quantum fluctuations—absent in theoretical predictions—is likely driven by the unique topology of the present ladder, where spin-1/2 and spin-5/2 chains, each expected to host TLL ground state, may couple synergistically to promote classical order.
This system therefore serves as a valuable benchmark for quantifying how quantum critical states, such as the TLL, become destabilized by classical correlations and lattice topology.

In summary, we synthesized a radical-Mn polymer, [Mn($p$-Py-V)$_2$(NO$_3$)$_2$]$_n$, that realizes a spin-$(1/2,5/2)$ three-leg ladder.
The magnetization curves and ESR modes are well explained by classical MF calculations. 
Furthermore, comparison with QMC simulations shows that the quantum fluctuations, expected to dominate in the present ladder, are entirely quenched in the real system, indicating a striking suppression of quantum effects.
Our findings provide a key experimental benchmark for the quantum–classical crossover and suggest that the ladder topology plays an essential role in shaping the balance between quantum and classical correlations in strongly correlated systems.

We thank K. Okunishi and H. Kamebuchi for valuable discussions.
This research was partly supported by the Shorai Foundation for Science and Technology.
A part of this work was performed under the interuniversity cooperative research program of the joint-research program of ISSP, the University of Tokyo.

The data that support the findings of this article are openly available~\cite{ripo}.


\end{document}